\documentclass[conference]{IEEEtran}

\IEEEoverridecommandlockouts

\usepackage[fleqn]{amsmath}
\usepackage{amssymb,amsfonts}
\usepackage{algorithmic}
\usepackage{graphicx}
\usepackage{textcomp}
\usepackage{xcolor}
\usepackage{mathtools}
\usepackage{longtable}
\usepackage{caption}
\usepackage{float}
\usepackage{url}

\DeclareMathSizes{6.5}{6.5}{6.5}{6.5}
\usepackage{multicol}
\usepackage{multirow}
\usepackage{newfloat}
\usepackage{enumerate}
\usepackage{framed}
\usepackage[lined,boxed,commentsnumbered]{algorithm2e}
\SetAlCapSkip{1em}

\usepackage{pgfplots}
\pgfplotsset{compat=1.14}
\usepackage{adjustbox}
\usepackage{blindtext}
\usepackage{tikz}
\usepackage{cleveref}
\usetikzlibrary{positioning}
\SetKwInOut{Parameter}{parameter}
\SetKw{KwBy}{by}
\newcommand{\argmin}{\mathop{\mathrm{argmin}}}
\def\BibTeX{{\rm B\kern-.05em{\sc i\kern-.025em b}\kern-.08em
    T\kern-.1667em\lower.7ex\hbox{E}\kern-.125emX}}
    
\usepackage{filecontents}
\usepackage[noadjust]{cite}

\begin{document}

\title{Secure and Efficient Federated Transfer Learning\\
}

\author{\IEEEauthorblockN{Shreya Sharma\IEEEauthorrefmark{1}, Chaoping Xing\IEEEauthorrefmark{2}, Yang Liu\IEEEauthorrefmark{3} and Yan Kang\IEEEauthorrefmark{3}} 
\IEEEauthorblockA{\IEEEauthorrefmark{1}Department of Electronics Engineering, 
Indian Institute of Technology (BHU) Varanasi, India\\
Email: shreyas.cd.ece17@iitbhu.ac.in} \IEEEauthorblockA{\IEEEauthorrefmark{2}School of Electronics, Information \& Electrical Engineering, 
Shanghai Jiao Tong University, China \\
School of Physical \& Mathematical Sciences, 
Nanyang Technological University, Singapore\\
Email: xingcp@ntu.edu.sg}
\IEEEauthorblockA{\IEEEauthorrefmark{3}Webank, Shenzhen, China\\ 
Email: yangliu@webank.com, yangkang@webank.com}}

\maketitle

\begin{abstract}
\begin{quote}
Machine Learning models require a vast amount of data for accurate training. In reality, most data is scattered across different organizations and cannot be easily integrated under many legal and practical constraints.
Federated Transfer Learning (FTL) was introduced in \cite{SecureFTL} to improve statistical models under a data federation that allow knowledge to be shared without compromising user privacy, and enable complementary knowledge to be transferred in the network. As a result, a target-domain party can build more flexible and powerful models by leveraging rich labels from a source-domain party. However, the excessive computational overhead of the security protocol involved in this model rendered it impractical.

In this work, we aim towards enhancing the efficiency and security of existing models for practical collaborative training under a data federation by incorporating Secret Sharing (SS). In literature, only the semi-honest model for Federated Transfer Learning has been considered. In this paper, we improve upon the previous solution, and also allow malicious players who can arbitrarily deviate from the protocol in our FTL model. This is much stronger  than  the semi-honest model where we assume that parties follow the protocol precisely. We do so using the one of the practical MPC protocol called SPDZ, thus our model can be efficiently extended to any number of parties even in the case of a dishonest majority.

In addition, the models evaluated in our setting significantly outperform the previous work, in terms of both runtime and communication cost. A single iteration in our model executes in 0.8 seconds for the semi-honest case and 1.4 seconds for the malicious case for 500 samples, as compared to 35 seconds taken by the previous implementation. 
\end{quote}
\end{abstract}

\section{Introduction}
The application of Artificial Intelligence (AI) is driven by big data availability. To calibrate the performance of real-life complex models such as AlexNet~\cite{AlexNet} with 60 million parameters and 650,000 neurons, large datasets providing millions of samples are used. AlphaGo~\cite{AlphaGo}, was trained on a collection of 29.4 million moves from 160,000 actual games.  However, high quality data is not readily available to meet such extensive requirements. Across various organisations, data is present in small quantities (i.e. few samples and labels) and is not heavily supervised (i.e. exists in unlabeled form). Thus, data needed for a particular task might not be present in a single place. Such a scenario can be dealt with by the unification of datasets from different platforms for better statistical modelling. But this approach lacks practicality due to the privacy concerns involved. The datasets available might contain sensitive information such as medical-records or financial data. Strict legislative laws that require explicit user approval before data usage further bound the merging of data. Apart from this, any form of well-supervised (labelled) data might require confidentiality because it constitutes a monetary or competitive advantage for the parties that own it. Thus arises the need for AI models that comply with both security and accuracy.

One solution to the above problem can be Federated Learning~\cite{FLGoogle}. This system enables parties to learn a shared prediction model while keeping all the training data locally stored. A prerequisite for this framework is data being horizontally partitioned (i.e. sharing a common feature space). As an alternative, secure machine learning on vertically partitioned data (i.e. data partitioned in the feature space) has also been studied in depth \cite{vertical1, vertical3}. Since these approaches require data to either have common samples or common features, they leave majority of the non-overlapping data underutilized. 

In this paper, we work on Federated Transfer Learning, where the target-domain party builds a prediction model by leveraging rich labels from a source-domain. This framework provides results for the entire feature and sample space as it doesn't place any restrictions on the distribution of data and thus can find applicability in diverse fields. For instance, many businesses can rely their decisions on the financial activities of different customers and since banks own authentic labels on such financial activities, a collaborative model trained using the datasets of two such organizations can benefit such businesses. In this case, due to involvement of different institutions for training, a small portion of the feature space will overlap, and only a part of the entire user space would intersect. Thus, here FTL becomes an important extension of Federated Learning. Additional security protocols are needed to maintain privacy in the case of FTL as the training involves collaborative calculations using the parameters related to data of both source and target domain. Although our work focusses on a two-party case, we discuss the importance of its scalability in the last section of the paper.

\subsection{Our Contribution}
Our work takes the FTL model introduced in \cite{SecureFTL}, and:
\begin{itemize}
    \item Enhances the efficiency by an order of magnitude in the semi-honest security setting by implementing the model using Multi-Party Computation (MPC) as opposed to the previous Homomorphic Encryption (HE) based approach.  
    \item Introduces active security to it, dealing with the case where $m-1$ out of the $m$ parties can deviate from the protocol. We do so using the SPDZ protocol\cite{SPDZ}, thus our model can be efficiently extended to any number of parties. To the best of our knowledge, ours is the first work to achieve active security for any non-linear deep learning based model for a dishonest majority.
    \item We present extensive experimental results that underscore the practicality of secure ML even in the active setting. We investigate different aspects related to scalability of the model and also highlights a comparison between the performance in our setting versus the previous work on secure FTL in the semi-honest setting. Our model significantly outperforms this HE-based implementation in terms of both communication cost and runtime.
\end{itemize}

\subsection{Related Work}
Federated Learning can be seen as decentralized machine learning, and thus is closely related to multi-party privacy preserving machine learning. This field has undergone extensive research in recent years due to its vast applicability. CryptoNets \cite{CryptoNets} based on Leveled Homomorphic Encryption (LHE), enabled encrypted prediction on a server-side model. But the use of LHE results in high computational overhead and since LHE restricts the degree of the polynomial used in the approximation of non-linear activation functions, the model wasn't accurate enough. SecureML \cite{SecureML} focuses on private training and inference of various ML models involving multiple parties. The approach provides security in the semi-honest model and incurred a loss of accuracy due to the crypto-friendly techniques involved. For the evaluation of certain non-linear activation functions like softmax, it proposes an expensive switch between arithmetic secret shares and Garbled circuits. Recent state-of-the-art work in the realm of privacy preservation in ML do so by outsourcing computation to a non-colluding server, and fail to achieve active security in the case of a dishonest majority. 

There aren't many instances of work on actively secure ML, as this setting is far less efficient as compared to its semi-honest counter parts. \cite{Helen} showed promising results for ML training in an actively secure setting, but this framework is only applicable for linear regularized models. \cite{valerie} is another work in the active setting that shows linear regression and logistic regression using SPDZ can match similar latency as SecureML.

All the frameworks so far do not consider the cases where data sets are small or have weak supervision. To address this scenario, secure Federated Transfer Learning was explored in \cite{SecureFTL}. It was the first framework enabling federated learning to benefit from transfer learning in a secure way. The work presented experimental results for a model based on HE and provided security in the semi-honest setting, with the major drawback being a lack of efficiency.

\section{Problem Definition}\label{problem}
Consider $\mathcal{D}_S$ and $\mathcal{D}_T$ are two datasets owned by two different parties and need to be kept private. The label rich source domain dataset can be defined as $\mathcal{D}_S \coloneqq \{(x_i^S, y_i^S)\}_{i=1}^{N_S}$, where $x_i^S \in \mathbb{R}^d$ is the $i^{th}$ sample and $y_i^S \in \{0,1\}$ is the $i^{th}$ label. And the target domain dataset can be defined as $\mathcal{D}_T \coloneqq \{(x_j^T\}_{j=1}^{N_T}$ where $x_j^T \in \mathbb{R}^d$ is the $j^{th}$ sample. We work on the case where there is a co-occurrence of certain samples i.e. $\mathcal{D}_{ST} \coloneqq \{(x_k^{ST}, x_k^{ST})\}_{k=1}^{N_{ST}}$ and, certain samples in $\mathcal{D}_T$ have a label in $\mathcal{D}_S$ i.e. $\mathcal{D}_L \coloneqq \{(x_k^L, y_k^L)\}_{k=1}^{N_L}$. The sample IDs are masked with an encryption scheme which enables the sharing of set of common IDs. We assume that both parties already know the commonly shared IDs.

In this paper, we aim to enable the two parties to build a transfer learning model to predict accurate labels for the target domain dataset while keeping their data private against an adversary. We work on two threat models i.e. \textit{semi-honest} and \textit{malicious}. In the former, an adversary can corrupt one of the two parties, and try to know more information about the private data of the other party than what can be inferred from the output, while following the protocol. In the latter, the adversary can corrupt one of the parties and make it deviate randomly from the protocol specification in order to falsify the output or learn private data of the other party. In the malicious case, we assume that for every iteration the local computations by the adversary on its own data are honest, and it tries to cheat during the interactive calculations on the joint data of the parties. This assumption is reasonable since any MPC protocol guarantees security against an adversary trying to know additional information from the visible messages during protocol execution and not for the case where the protocol starts with false inputs. 

\section{Preliminaries} \label{prelim}
(Additive) Secret Sharing protocols rely on splitting every private value involved in the computation (of arithmetic circuits) into additive secret shares. Such a sharing enables linear operations (i.e. addition and scalar multiplication) on the actual values, to be performed locally (without interaction) on the shares in order to obtain the corresponding shares of the output. In contrast, each multiplication of shared values requires a unique multiplication triple \cite{Bea91}. These triples are input-independent and can be generated at any time before the execution of the protocol. Thus these protocols are executed in two phases, the computationally expensive offline phase and a relatively cheap (information-theoretic) online phase. The offline phase is run to generate raw material, which is consumed later by the online phase.

We implement the secret sharing component of our protocol in the ABY framework \cite{ABY} for the semi-honest setting, and in the SPDZ framework \cite{MP-SPDZ} for the malicious setting. In this section, we summarize these two frameworks. 

\subsection{ABY}
ABY is a mixed-protocol framework that combines various secret sharing schemes in a two-party setting, namely Arithmetic, Boolean and Yao Sharing. Our work is based on Arithmetic Sharing i.e. the values are additively shared in a ring $\mathbb{Z}_{2^l}$, in a semi-honest threat model. The offline phase in this framework involves generation of multiplication triples using Oblivious Transfer (OT).  An efficient way to perform many OTs is to extend a small number of expensive baseOTs (based on asymmetric cryptographic operations) using OTExtension (based on much faster symmetric cryptographic primitives) in a constant number of rounds \cite{OTExtension}. The online protocol for arithmetic sharing in ABY is described in Figure \ref{ABY}. 
\begin{figure}[!htb]
\caption{$\Pi_{\mathsf{Online}}^{\mathsf{ABY}}$ - ABY Online Protocol}
\begin{framed}
\textbf{Input:} To input a value $x$, Party $P_i$ chooses $r \in \mathbb{Z}_{2^l}$, sets its share $\langle x \rangle_i = x - r$ and sends $r$ to $P_{1-i}$, who sets $\langle x \rangle_{1-i
}= r$. \\
\textbf{Add:} $\langle z \rangle = \langle x \rangle + \langle y \rangle$: $P_i$ locally computes $\langle z \rangle_i = \langle x \rangle_i + \langle y \rangle_i$. \\
\textbf{Multiply:} $\langle z \rangle = \langle x \rangle \cdot \langle y \rangle$ : a multiplication triple generated of the form $\langle c \rangle = \langle a \rangle \cdot \langle b \rangle$ during offline phase is taken. $P_i$ sets $\langle \epsilon \rangle_i = \langle x \rangle_i - \langle a \rangle_i$ and $\langle \rho \rangle_i = \langle y \rangle_i - \langle b \rangle_i$. Parties output $\epsilon$ and $\rho$. $P_i$ sets $\langle z \rangle_i = i \cdot \epsilon \cdot \rho + \rho \cdot \langle a \rangle_i + \epsilon \cdot \langle b \rangle_i + \langle c \rangle_i$ \\
\textbf{Output:} To output a value $x$ to $P_i$, $P_{1-i}$ sends its share $\langle x \rangle_{1-i}$ to $P_{i}$ who computes  $x = \langle x \rangle_0 + \langle x \rangle_1$.
\end{framed}
\label{ABY}
\end{figure}

\begin{figure}[!htb]
\caption{$\Pi_{\mathsf{Online}}^{\mathsf{SPDZ}}$ - SPDZ Online Protocol}
\begin{framed}
The set ${P}$ is the complete set of parties.
\\
\textbf{Initialise:} The parties call preprocessing functionality to obtain enough multiplication triples $(\langle a \rangle, \langle b \rangle, \langle c \rangle)$ and input mask values $(r_j, \langle r_j \rangle)$ according to the function being evaluated. If the functionality aborts, the parties output $\perp$
and abort.
\\
\textbf{Input:} To input $x_j$, party ${P}_j\in{P}$ takes a mask value $(r_j, \langle r_j \rangle)$, then:
\begin{enumerate}
\setlength\itemsep{1pt}
\item Broadcasts $\Delta \leftarrow x_j - r_j$

\item Parties compute $\langle x_j \rangle \leftarrow \langle r_j \rangle + \Delta$
\end{enumerate}

\textbf{Add:} On input $(\langle x \rangle, \langle y \rangle)$, locally compute $\langle x+y \rangle \leftarrow \langle x \rangle + \langle y \rangle$
\\
\textbf{Multiply:} On input $(\langle x \rangle, \langle y \rangle)$, the parties:
\begin{enumerate}
\setlength\itemsep{1pt}
\item Take a multiplication triple $(\langle a \rangle, \langle b \rangle, \langle c \rangle)$, compute $\langle \epsilon \rangle \leftarrow \langle x \rangle - \langle a \rangle$ and $\langle \rho \rangle \leftarrow \langle y \rangle - \langle b \rangle$ and partially open them to obtain $\epsilon$ and $\rho$.

Partially opening a share involves each party sending its own share of the value to every other party and computing the sum of all the shares available to it, while the corresponding MAC value $\gamma(x_i)$ is kept secret.

\item Set $\langle z \rangle \leftarrow \epsilon \cdot \rho$ + $\epsilon \cdot \langle a \rangle$ + $\rho \cdot \langle b \rangle$ + $\langle c \rangle$. 
\end{enumerate}

\textbf{Output:} To output a share $\langle x \rangle$:
\begin{enumerate}
\setlength\itemsep{1pt}
\item Check all partially opened values since the last batched MAC-check, as follows:
\begin{itemize}
\setlength\itemsep{2pt}
\item The parties have ids $id_1$, ... $id_k$ corresponding to opened values $x_1$, .. $x_k$

\item Parties agree on a random vector \small{$\mathbf{r} \leftarrow  \mathcal{F}_{\mathrm{Rand}}\left(\mathbb{F}_{q}^{k}\right)$}

\item Party ${P}_i$ computes $c \leftarrow \sum_{j=1}^{k} r_{j} \cdot x_{j}$ and $\gamma(c)_{i} \leftarrow \sum_{j=1}^{k} r_{j} \cdot \gamma\left(x_{j}\right)_{i}$

\item Parties run batched MAC-check on $c$, where party ${P}_i$ inputs $c$ and $\gamma(c)_i$.

\end{itemize}

\item If the MAC-check fails, output $\perp$ and abort.

\item Open each party ${P}_i$'s input sent to every other party ${P}_j$, to compute $x \leftarrow \sum_{i \in {P}} x_{i}$. Run MAC-check with party ${P}_i$'s input $x$ and $\gamma(x_i)$, to verify $\langle y \rangle$. In case this check fails, output $\perp$ and abort; otherwise output $x$.
\end{enumerate}
\end{framed}
\label{SPDZ}
\end{figure}

\subsection{SPDZ}
SPDZ is a family of Multi-Party Computation (MPC) protocols for arbitrary number of parties, providing active security with information theoretic MACs, in the case where majority of the parties are corrupted. More specifically, every party $P_i$ has an additive share of the fixed MAC key i.e. $\alpha_i \in \mathbb{F}_p$ such that $\alpha = \alpha_1 + ... + \alpha_n$. A data-item is $\langle\cdot\rangle$-shared, if every Party $P_i$ has a tuple $(a_i, \gamma(a)_i)$ such that $a_i$ is an additive share of $a$ i.e. $a = a_1 + ... a_n$ and $\gamma(a)_i$ is an additive share of the corresponding MAC $\gamma(a)$ i.e. $\gamma(a) = \alpha a$ and $\gamma(a) = \gamma(a)_1 + .. + \gamma(a)_n$. The online-phase of the SPDZ protocol is given in  Figure~\ref{SPDZ}. 

\section{Secure FTL Interface}\label{interface}
In this section, we explain the FTL model used for our work and then present an algorithm to train this model using Secret Sharing. 

We assume that the parties S and T have produced a hidden representation of their data using neural networks $Net^T$ and $Net^S$ i.e. $u_i^S$ = $Net^S(x_i^S)$ and $u_i^T$ = $Net^T(x_i^T)$, where $u^S \in \mathbb{R}^{N_S \times d}$ and $u^T  \in \mathbb{R}^{N_T \times d}$, $d$ is the dimension of hidden layer. In order to generate labels for target domain the following translator function from \cite{equation} is used: 
\begin{center}
$
\psi(u_j^T) = \frac{1}{N_S} \sum_i^{N_S} y_i^S \cdot u_i^S \cdot (u_j^T)^\prime
$
\end{center}
For simplification, translator function is seen to be of the form, $\psi(u_j^T) = \Lambda^S \mathcal{C}(u_j^T)$, where $\Lambda^S = \frac{1}{N_S} \sum_i^{N_S} y_i^S \cdot u_i^S $ and $\mathcal{C}(u_j^T)=(u_j^T)^\prime$.

Then for training purpose, we follow:
\begin{center}
$
    \underset{\omega^S, \omega^T} {\operatorname{argmin}} $ $ \mathcal{L}_1 = \sum^{N_L}_{i=1} \ell_1(y_i^S, \psi(u_i^T))
$
\end{center}
In the above equation, $\omega^S$ and $\omega^T$ are training parameters of $Net^S$ and $Net^T$ respectively. If $Net^S$ has $l_S$ layers, $\omega^S$ = $\{\omega_l^S\}_{l=1}^{l_S}$, similarly if $Net^T$ has $l_T$ layers, $\omega^T$ = $\{\omega_l^T\}_{l=1}^{l_T}$, where $\omega_l^S$ and $\omega_l^T$ are training parameters for the $l^{th}$ layer. $\ell_1$ is loss function used i.e. $\ell_{1}(y, \psi)=\log (1+\exp (-y \psi))$.
In order to achieve feature transfer learning in federated learning setting, we follow:
\begin{center}
$
\underset{\omega^S, \omega^T} {\operatorname{argmin}} $ $\mathcal{L}_{2}=-\sum_{i}^{N_{ST}} \ell_{2}\left(u_{i}^{S}, u_{i}^{T}\right)
$,
\end{center}
where $\ell_2$ is alignment loss, i.e. $\ell_{2}\left(u_{i}^{S}, u_{i}^{T}\right) = \kappa u_{i}^{S}\left(u_{i}^{T}\right)^{\prime}$ and $\kappa=-1$. 

By combining the two loss equations with regularization terms, the objective of training becomes:
\begin{equation}
\label{lossnet}
\argmin_{\omega^S, \omega^T}  \mathcal{L}=\mathcal{L}_{1}+\gamma \mathcal{L}_{2}+\frac{\lambda}{2}\left(\mathcal{L}_{3}^{S}+\mathcal{L}_{3}^{T}\right),
\end{equation}
where the weight parameters are given by $\lambda$ and $\gamma$, and the regularization terms are given as, $\mathcal{L}_3^S = \sum_{l}^{l_{S}}\left\|\omega_{l}^{S}\right\|_{F}^{2}$ and $\mathcal{L}_{3}^{T}=\sum_{l}^{l_{T}}\left\|\omega_{l}^{T}\right\|_{F}^{2}$.\\

The relation followed for back-propagation:
\begin{equation}
\label{backprop}
\frac{\partial \mathcal{L}}{\partial \omega_{l}^{i}}=\frac{\partial \mathcal{L}_{1}}{\partial \omega_{l}^{i}}+\gamma \frac{\partial \mathcal{L}_{2}}{\partial \omega_{l}^{i}}+\lambda \omega_{l}^{i}.
\end{equation}

In order to prevent leakage of any data for both parties A and B, we use secret sharing to collaboratively compute Eq. \ref{lossnet} and Eq. \ref{backprop}. For better suitability to the framework used, we use second order Taylor approximation for the $\ell_1$ function  i.e.
\begin{equation}
    \ell_{1}(y, \psi) \approx \ell_{1}(y, 0)+\frac{1}{2} \mathcal{T}_1(y) \psi+\frac{1}{8} \mathcal{T}_2(y) \psi^{2}.
\end{equation}

where, $\mathcal{T}_1(y)=\left.\frac{\partial \ell_{1}}{\partial \psi}\right|_{\psi=0}$, $\mathcal{T}_2(y)=\left.\frac{\partial^{2} \ell_{1}}{\partial \psi^{2}}\right|_{\psi=0}$.

\begin{equation}
\frac{\partial \ell}{\partial \psi}=\frac{1}{2} \mathcal{T}_1(y)+\frac{1}{4} \mathcal{T}_2(y) \psi
\end{equation}
For logistic loss, $\mathcal{T}_1(y)=y$, $\mathcal{T}_2(y)=y^{2}$.
This approximation helps avoid the use of expensive techniques for division by a secret value, as required in the calculation of (exp$(-y\psi)$) for $\ell_1$.

Hence the final loss and gradient equations become:
\begin{equation}
\begin{aligned}\mathcal{L} &=\sum_{i}^{N_{L}}\bigg(\ell_{1}\left(y_{i}^{S}, 0\right)+\frac{1}{2} \mathcal{T}_1\left(y_{i}^{S}\right) \Lambda^{S}\mathcal{C}\left(u_{i}^{T}\right)\\ &+\frac{1}{8} \mathcal{T}_2\left(y_{i}^{S}\right) \Lambda^{S}\mathcal{C}\left(u_{i}^{T}\right) \Big(\Lambda^{S}\mathcal{C}\left(u_{i}^{T}\right)\Big) \bigg) \\ &+\gamma \sum_{i}^{N_{S T}}\bigg(\ell_{2}^{T}(u_{i}^{T})+\ell_{2}^{S}\left(u_{i}^{S}\right)+\kappa u_{i}^{S}\left(u_{i}^{T}\right)^{\prime}\bigg) \\ &+\frac{\lambda}{2} \mathcal{L}_{3}^{S}+\frac{\lambda}{2} \mathcal{L}_{3}^{T} \end{aligned}
\label{loss}
\end{equation}
\begin{equation}
\begin{aligned}\frac{\partial \mathcal{L}}{\partial \omega_{l}^{T}} &=\sum_{i}^{N_{L}} \bigg(\frac{1}{4} \mathcal{T}_2\left(y_{i}^{S}\right)\Lambda^{S}\mathcal{C}\left(u_{i}^{T}\right) \Big(\Lambda^S \frac{\partial \mathcal{C}\left(u_{i}^{T}\right)}{\partial \omega_{l}^{T}}\Big)\\ &+\frac{1}{2} \mathcal{T}_1\left(y_{i}^{S}\right) \Lambda^{S} \frac{\partial \mathcal{C}\left(u_{i}^{T}\right)}{\partial \omega_{l}^{T}}\bigg) \\ &+\sum_{i}^{N_{ST}}\left(\gamma \kappa u_{i}^{S}\frac{\partial u_{i}^{T}}{\partial \omega_{l}^{T}}+\gamma \frac{\partial \ell_{2}^{T}\left(u_{i}^{T}\right)}{\partial \omega_{l}^{T}}\right)+\lambda \omega_{l}^{T} \end{aligned}
\label{gradient_b}
\end{equation}
\begin{equation}
\begin{aligned}\frac{\partial \mathcal{L}}{\partial \omega_{l}^{S}} &= \sum_{i}^{N_{L}}\bigg(\frac{1}{4} \mathcal{T}_2\left(y_{i}^{S}\right) \Lambda^{S}\mathcal{C}(u_{i}^{T})\Big(\frac{\partial \Lambda^S}{\partial \omega_{l}^{S}} \mathcal{C}\left(u_{i}^{T}\right)\Big)\\ &+\frac{1}{2} \mathcal{T}_1\left(y_{i}^{S}\right)\mathcal{C}\left(u_{i}^{T}\right) \frac{\partial \Lambda^{S}}{\partial \omega_{l}^{S}} \bigg) \\ &+\gamma \sum_{i}^{N_{ST}}\left(\kappa u_{i}^{T} \frac{\partial u_{i}^{S}}{\partial \omega_{l}^{S}}+\frac{\partial \ell_{2}^{S}\left(u_{i}^{S}\right)}{\partial \omega_{l}^{S}}\right)+\lambda \omega_{l}^{S} \end{aligned}
\label{gradient_a}
\end{equation}

The components of the Eq. \ref{loss}, \ref{gradient_b}, \ref{gradient_a} can be divided into three categories according to the ones that can be locally computed by S and T, and the ones that are to be computed on the joint data of both parties.  

For Eq. \ref{loss}:
\begin{equation}
\begin{aligned}\mathcal{L}_1^{ST} = \sum_{i}^{N_{L}} \frac{1}{8} \mathcal{T}_2\left(y_{i}^{S}\right) \Lambda^{S}\mathcal{C}\left(u_{i}^{T}\right) \Big(\Lambda^{S}\mathcal{C}\left(u_{i}^{T}\right)\Big) \\+ \frac{1}{2} \mathcal{T}_1\left(y_{i}^{S}\right) \Lambda^{S}\mathcal{C}\left(u_{i}^{T}\right) + \sum_{i}^{N_{ST}} \kappa u_{i}^{S}\left(u_{i}^{T}\right)^{\prime} \end{aligned}
\end{equation}
\begin{equation}
\mathcal{L}_1^S = \sum_{i}^{N_{L}} \ell_{1}\left(y_{i}^{S}, 0\right) + \sum_{i}^{N_{ST}} \ell_{2}^{S}\left(u_{i}^{S}\right) + \frac{\lambda}{2} \mathcal{L}_{3}^{S} 
\end{equation}
\begin{equation}
\mathcal{L}_1^T = \gamma \sum_{i}^{N_{ST }}\ell_{2}^{T}(u_{i}^{T}) + \frac{\lambda}{2} \mathcal{L}_{3}^{T} 
\end{equation}

For Eq. \ref{gradient_b}:
\begin{equation}
\begin{aligned}
\mathcal{L}_2^{ST} = \sum_{i}^{N_{L}} \frac{1}{4} \mathcal{T}_2\left(y_{i}^{S}\right)\Lambda^{S}\mathcal{C}\left(u_{i}^{T}\right) \Big(\Lambda^S \frac{\partial \mathcal{C}\left(u_{i}^{T}\right)}{\partial \omega_{l}^{T}}\Big) + \\\frac{1}{2} \mathcal{T}_1\left(y_{i}^{S}\right) \Lambda^{S} \frac{\partial \mathcal{C}\left(u_{i}^{T}\right)}{\partial \omega_{l}^{T}} +\sum_{i}^{N_{ST}}\left(\gamma \kappa u_{i}^{S}\frac{\partial u_{i}^{T}}{\partial \omega_{l}^{T}}\right)
\end{aligned}
\end{equation}
\begin{equation}
\begin{aligned}\mathcal{L}_2^T &= \sum_{i}^{N_{ST}}\gamma \frac{\partial \ell_{2}^{T}\left(u_{i}^{T}\right)}{\partial \omega_{l}^{T}})+\lambda \omega_{l}^{T}
\end{aligned}
\end{equation}

For Eq. \ref{gradient_a}:
\begin{equation}
\begin{aligned}
\mathcal{L}_3^{ST} &= \sum_{i}^{N_{L}}\bigg(\frac{1}{4} \mathcal{T}_2\left(y_{i}^{S}\right) \Lambda^{S}\mathcal{C}(u_{i}^{T})\Big(\frac{\partial \Lambda^S}{\partial \omega_{l}^{S}} \mathcal{C}\left(u_{i}^{T}\right)\Big)\\ &+\frac{1}{2} \mathcal{T}_1\left(y_{i}^{S}\right)\mathcal{C}\left(u_{i}^{T}\right) \frac{\partial \Lambda^{S}}{\partial \omega_{l}^{S}} \bigg) + \gamma \sum_{i}^{N_{ST}} \kappa u_{i}^{T} \frac{\partial u_{i}^{S}}{\partial \omega_{l}^{S}} 
\end{aligned}
\end{equation}
\begin{equation}
\begin{aligned}
\mathcal{L}_3^S &= \gamma \sum_{i}^{N_{ST}}\left(\frac{\partial \ell_{2}^{S}\left(u_{i}^{S}\right)}{\partial \omega_{l}^{S}}\right)+\lambda \omega_{l}^{S}
\end{aligned}
\end{equation}

Using the notations stated above, we construct Algorithm~1 to execute the FTL training, where $\Pi^\mathsf{SS}_\mathsf{Online}$ can be $\Pi^\mathsf{ABY}_\mathsf{Online}$ or $\Pi^\mathsf{SPDZ}_\mathsf{Online}$. Once the model has been trained, party T can use it to obtain predictions for its data samples. This would involve the parties computing $\psi(x^T)$ collaboratively, and then S sending the predicted label for this entry by the federated model to T.

Although the collaborative computations are secured by the secret-sharing schemes we use, our protocol does involve revealing some values after each iteration, which would have compromised privacy. To combat any corrupt move based on the information attained from this part of the training, we reveal whether {$\mathcal{L}_{prev}-\mathcal{L} < \epsilon$} follows or not, rather than revealing the value of $\mathcal{L}$ itself. This comparison can performed in a secure way on the shares itself in both our frameworks, without having to reveal the actual value. Moreover, we reveal the respective gradients of the parties only to them, thereby ensuring privacy. 
\newline
\textbf{Theorem 1: } \textit{Algorithm 1 using $\Pi^\mathsf{ABY}_\mathsf{Online}$ is information theoretically secure against a semi-honest adversary.} \\
\textit{Proof} ABY guarantees that no additional information is revealed except for the outputs. The loss function being revealed after each iteration is masked and the gradients of each party are only revealed to that particular party, thereby maintaining privacy. \\
\newline
\textbf{Theorem 2:} \textit{Algorithm 1 using $\Pi^\mathsf{SPDZ}_\mathsf{Online}$ for interactive calculation is information theoretically secure against a malicious adversary.} \\
\textit{Proof} Given that the SPDZ protocol is secure against active adversaries in the two-party setting, the proof for active security follows exactly like that of Theorem 1.

\begin{algorithm}[!htb]
\SetAlgoLined 
\SetKwInOut{Input}{Input}\SetKwInOut{Output}{Output}
\Input{learning rate $\eta$, weight parameter $\gamma$, $\lambda$, max iteration $m$, tolerance $\epsilon$}
\Output{ Model parameters $\omega^S, \omega^T$} 
\BlankLine
S, T initialize $\omega^S, \omega^T$ \\
$iter = 0$\; \While{iter $\leq$ m}{
\textbf{S} do:\\
$u_i^S \leftarrow Net^S(\omega^S, x_i^S)$ for $i \in \mathcal{D}_S$ \;
\textbf{T} do: \\
$u_i^T \leftarrow Net^T(\omega^T, x_i^T)$ for $i \in \mathcal{D}_T$ \;
\textbf{S, T} do: \\
Obtain $\langle \mathcal{L}^{ST}_1 \rangle_k$ using $\Pi_\mathsf{Online}^{\mathsf{SS}}$, where $k$ = \{S, T\} \;
\textbf{S, T} do: \\
Set $\langle \mathcal{L}\rangle_k$ = $\mathcal{L}_1^{k}$ + $\langle \mathcal{L}^{ST}_1 \rangle_k$, where $k$ = \{S, T\} \;
\For{$l\gets0$ \KwTo $l_S$ \KwBy $1$}{
\textbf{S, T} do: \\
Obtain $\langle \mathcal{L}^{ST}_2 \rangle_k$ using $\Pi_\mathsf{Online}^{\mathsf{SS}}$, where $k$ = \{S, T\} \;
\textbf{S, T} do: \\
Set $\Big \langle \frac{\partial \mathcal{L}}{\partial \omega_{l}^{T}} \Big \rangle_S $ = $\langle \mathcal{L}^{ST}_2 \rangle_S$ and $\Big \langle \frac{\partial \mathcal{L}}{\partial \omega_{l}^{T}} \Big \rangle_T$ = $\langle \mathcal{L}^{ST}_2 \rangle_T$ + $\mathcal{L}_2^T$. Output the value $ \frac{\partial \mathcal{L}}{\partial \omega_{l}^{T}} $ to $T$ \;
\textbf{T} do: \\
Update $\omega^T_l$ \; 
}
\For{$l\gets0$ \KwTo $l_T$ \KwBy $1$}{
\textbf{S, T} do: \\
Obtain $\langle \mathcal{L}^{ST}_3 \rangle _k$ using $\Pi_\mathsf{Online}^{\mathsf{SS}}$, where $k$ = \{S, T\} \; 
    \textbf{S, T} do: \\
Set $\Big \langle \frac{\partial \mathcal{L}}{\partial \omega_{l}^{S}} \Big \rangle_T $ = $\langle \mathcal{L}^{ST}_3 \rangle_S$ and $\Big \langle \frac{\partial \mathcal{L}}{\partial \omega_{l}^{S}} \Big \rangle_S$ = $\langle \mathcal{L}^{ST}_3 \rangle_S$ + $\mathcal{L}_3^S$. Output the value $ \frac{\partial \mathcal{L}}{\partial \omega_{l}^{S}} $ to $S$ \;
\textbf{T} do: \\
Update $\omega^S_l$ \;
}
Output the binary value $\mathcal{L}_{diff} = (\mathcal{L}_{prev}-\mathcal{L} < \epsilon)$ using $\Pi_\mathsf{Online}^{\mathsf{SS}}$ \;
\eIf{$\mathcal{L}_{diff} == 1$}{
    Stop the protocol \;
  }{
    Set $\langle \mathcal{L}_{prev}\rangle = \langle \mathcal{L} \rangle$ \;
    $iter \leftarrow iter + 1$ \;
  }
}
\caption{FTL Training}
\end{algorithm}

\section{Experiments}\label{evaluation}
\subsection{Setup}
We run all experiments by simulating both parties on a single Intel i5 machine with 16 GB memory. The MP-SPDZ Library~\cite{MP-SPDZ} is used to implement actively secure FTL training. The computation is conducted in a 64-bit prime field with statistical security parameter $\sigma= 40$. The offline phase of the protocol is based on LowGear version of Overdrive~\cite{Overdrive}, which is the most efficient preprocessing protocol for two parties. The online version is based on SPDZ-2~\cite{Cowgear}.
The semi-honest version of FTL training based on Secret Sharing is implemented using ABY Framework\cite{encrypto}. The ring size was chosen to be 64 bits i.e. $l = 64$, symmetrical security parameter to be $\kappa=128$ and $\sigma = 40$. Apart from this, the encrypted version introduced in \cite{SecureFTL} is emulated using FATE Framework \cite{FATE}. 

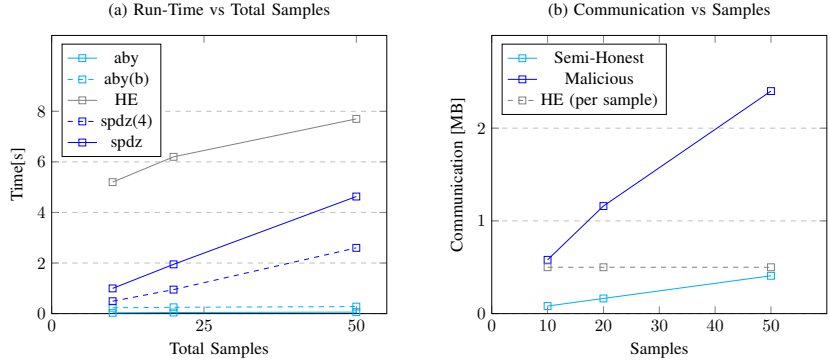
\begin{figure*}[!htb]
\centering

\begin{tikzpicture}[scale=0.65,>=latex]

\begin{axis}[
 title={(a) Run-Time vs Total Samples},
    xlabel={Total Samples},
    ylabel={Time[s]},
    xmin=0, xmax=55,
    ymin=0, ymax=11,
    xtick={0,25,50},
    ytick={0,2,4,6,8},
    legend pos=north west,
    ymajorgrids=true,
    grid style=dashed,
]

\addplot[
    color=cyan,
    mark=square,
    ]
    coordinates {
    (10,0.032)(20,0.043)(50,0.058)
    };
    \addlegendentry{aby}
\addplot[
    dashed,
    color=cyan,
    mark=square,
    mark options = solid,
    ]
    coordinates {
    (10,0.232)(20,0.250)(50,0.278)
    };
    \addlegendentry{aby(b)}
\addplot[
    color=gray,
    mark=square,
    ]
    coordinates {
    (10,5.2)(20,6.2)(50,7.7)
    };
    \addlegendentry{HE}   
\addplot[
    dashed,
    color=blue,
    mark=square,
    mark options = solid,
    ]
    coordinates {
    (10,0.49)(20,0.949)(50,2.6)
    };
    \addlegendentry{spdz(4)}
 \addplot[
    color=blue,
    mark=square,
    ]
    coordinates {
    (10,0.997)(20,1.948)(50,4.63)
    };
    \addlegendentry{spdz}    

\end{axis}
\begin{scope}[xshift=9.0cm]
\begin{axis}[
    title={(b) Communication vs Samples},
    xlabel={Samples},
    ylabel={Communication [MB]},
    xmin=0, xmax=60,
    ymin=0, ymax=3,
    xtick={0,10,20,30,40,50},
    ytick={0,1,2},
    legend pos=north west,
    ymajorgrids=true,
    grid style=dashed,
]
 
\addplot[
    color=cyan,
    mark=square,
    ]
    coordinates {
    (10,0.0817)(20,0.1635)(50,0.408)
    };
    \addlegendentry{Semi-Honest}

 \addplot[
    color=blue,
    mark=square,
    ]
    coordinates {
    (10,0.58)(20,1.16)(50,2.4)
    };
    \addlegendentry{Malicious}

\addplot[
    dashed,
    color=gray,
    mark=square,
    mark options = solid,
    ]
    coordinates {
    (10,0.5)(20,0.5)(50,0.5)
    };
    \addlegendentry{HE (per sample)}

\end{axis}
\end{scope}

\end{tikzpicture}
\captionsetup{font=footnotesize}
\caption{Comparison in Performance}
\label{fig:performance-plots}
\end{figure*}

\subsection{Model Specifications}
For our experiments related to scalability, we work on the Kaggle's Default-of-Credit-Card-Clients \cite{dataset} is used. This dataset contains credit card records of the users with the user's default payment as labels. The dataset offers 33 features and 30,000 samples after applying one-hot encoding to categorical features. To simulate the federation setting, the dataset is split in both samples and feature space. Party S, simulating the source domain, is given all the labels. Each sample is assigned to party T or S or both, wherein the co-occurring samples are termed as overlapping samples. For realistic experimentation we derive from our bank-business company example in the first section, the demographic features such as age, education etc. are provided to party T emulating business company, and features relating to financial activity such as bill balance data, six months of payment etc. to party S emulating the bank. 

For FTL Training, tests were conducted on NUS-WIDE dataset \cite{NUS}, which consists of Flickr images with 634 low-level image features and their associated 1000 tag features and 81 ground truth labels. One-vs-all classification  problem is considered with a data federation formed between party S and party T, where S has text tag features and image labels, and T has low-level image features. For  training  FTL, the three most  frequently occurring labels in the NUS-WIDE dataset are picked i.e., water, person and sky.  The training iterates until convergence or reaches the maximum iteration, i.e., 50, where $\gamma=0.05$ and $\lambda=0.005$. The local model we use for evaluation is a single layer stacked auto-encoder with domain invariant hidden layer representation $d=32$. 

\begin{table}
     \centering
     \captionsetup{font=footnotesize}
     \caption{Online-Phase runtime for varying sample size (Malicious)}
    \resizebox{0.65\linewidth}{!}{
     \begin{tabular}{|c|c|c|c|c|c|}
     \hline
           \multirow{2}{*}{Process} & \multicolumn{5}{c|}{Time[ms]} \\
          
           & 10 & 20 & 50 & 100 & 500\\

          \hline
          \hline
          
          Initialize  &5.3  & 8.7 & 19 & 44 &  205\\
          Computation  & 6.2 &  6.9 & 16 & 29 & 332\\
          Reveal &  15 &  22  & 36  & 53 &  292\\
          \hline
     \end{tabular}
     
     }
     \label{tab:spdz_online}
     
 \end{table}
 \begin{table}
     \centering
     \captionsetup{font=footnotesize}
     \caption{Runtime for varrying sample size (Semi-Honest)}
    \resizebox{\linewidth}{!}{
     \begin{tabular}{|c|c|c|c|c|c|c|}
     \hline
          \multirow{2}{*}{Phase} & Process & \multicolumn{5}{c|}{Time[ms]} \\
          & & 10 & 20 & 50 & 100 & 500\\
          \hline
          \hline
          Setup & BaseOT & 190.94  & 190.1& 194.7 & 201.4 & 196.3\\
          \hline
          \hline
          Offline & OTExtension & 38.18  & 55.7 & 106.1 &182.4 & 780.1\\
          \hline
          \hline
          Online &Computation & 4.1 & 7.8 & 21.55 & 44.3  & 220.6\\
          \hline
     \end{tabular}
     }
     \label{tab:aby}
     
 \end{table}    
\subsection{Results and Analysis}
Figure 3 compares the performance of different settings discussed for a single iteration, in terms of runtime and communication cost. Here \textit{aby} and \textit{aby(b)} stand for the runtime in semi-honest setting excluding and including baseOTs. Figure 4 further highlight results pertaining to the online SPDZ-based evaluation for a single iteration. All experiments were run on a single thread. Table \ref{accuracy} highlights the accuracy results comparing  SS-based  FTL  (SST) with the ones of HE-based FTL with Taylor loss (TLT) and FTL with logistic loss (TLL) in \cite{SecureFTL}.
Experimental results can be summarized as:\\[0.1cm]
\textsc{Comparison with HE-based implementation:}
\begin{itemize}
    \item[--] For 500 samples, the HE-based model takes around 35 seconds to evaluate where as our model takes 1.4 seconds in the malicious case and 0.8 seconds in the semi-honest case (excluding offline phase).
    \item[--] Even after including the time for preprocessing, our model outperforms the HE-based model, as evident in Figure 3.(a)
    \item[--] In contrast to the online communication cost for different sample sizes have been highlighted in Figure 3.(b), the HE-based model has a communication cost of about 0.50 MB per sample.
    \item[--] For the offline phase in malicious setting, generation of each triple requires sending 13.71 kbit of data, where using a single thread 8856 triples can be generated per second. The performance can be drastically improved by increasing the number of threads. The performance with 4 threads is highlighted in Figure 3.(a) labelled as \textit{spdz(4)}.
    \item[--] The offline phase of the semi-honest version sends $(\ell + 1)(\kappa + \ell)/2$ to generate an $\ell$ bit multiplication triple.
    \item[--] SS-based schemes can attain plain-text level accuracy whereas HE-based scheme incur a loss in accuracy, thus our model shows more accurate results on the NUS-WIDE dataset, as evident from Table \ref{accuracy}.

\end{itemize}

 \vspace{1pt}

\textsc{Secret Sharing based evaluation:}
\begin{itemize}
    \item[--] Table \ref{tab:spdz_online} and \ref{tab:aby} highlight that the runtime increases linearly with the increase in samples, since the number of inner products involved in the online phase is proportional to  $(aN_{ST}+bN_L)$, where $a$ and $b$ are two constants.
    \item[--] Revealing outputs is a bottleneck for small sample size in case of the SPDZ-based online evaluation (Table \ref{tab:spdz_online}). Since all the online communication stems from the multiplication gates evaluated, and the loss and gradient values revealed; for small sample sizes both the costs are comparable but as on increasing sample size the cost of evaluating multiplication gates shoots up and surpasses the cost of revealing. 
    \item[--] Figure 4.(a) suggests that on increasing features and overlapping samples, rate of increase in runtime becomes narrow i.e. training time will converge if we keep increasing features or overlapping samples. Figure 4.(b) highlights the quadratic growth in runtime of with respect to hidden layer dimension.
    \end{itemize}

\textsc{Scalablity SPDZ:}
 \begin{itemize}
     \item [--] We focus on a two-party case which is a highly favourable setting since instances of a two-party interaction are extensively available on the Internet in the form of client-server based protocols. Since this model requires domain-independent common representation layer, it is only applicable to a subset of transfer mechanisms. But the techniques use by us are flexible enough to implement any other model, and can be efficiently extended to arbitrary number of parties while guaranteeing strong security. 
     \item [--] Preprocessing throughput can further be increased upto 100000 triples per second by increasing the number of threads used to 16. 
     \item [--] Merging communication for multiple operations in a single round is the major reason behind the improvements achieved by this setting. Vectorizing more operations and increasing threads can lead to improvements in performance.
 \end{itemize}

 \begin{table}[ht]\Huge
\centering
\resizebox{0.45\textwidth}{!}{
    \begin{tabular}{|c|c|c c c|}\hline
    Tasks & $N_L$ & SST & TLT & TLL \\\hline
    \hline
    water vs. others &  $100$ & $\boldsymbol{0.698}\pm0.011$ & $0.692\pm0.062$ & $0.691\pm0.060$ \\
    water vs. others &  $200$ & $\boldsymbol{0.707}\pm0.013$ & $0.702\pm0.010$ & $0.701\pm 0.007$  \\
    person vs. others &  $100$ & $\boldsymbol{0.703}\pm0.015$ & $0.697\pm 0.010$ & $0.697 \pm 0.020$\\
    person vs. others & $200$ & $\boldsymbol{0.735}\pm0.004$ & $0.733 \pm 0.009$ & $\boldsymbol{0.735} \pm 0.010$ \\
    sky vs. others & $100$ & $0.708\pm0.015$ & $0.700\pm0.022$ & $\boldsymbol{0.713}\pm 0.006$ \\
    sky vs. others & $200$ & $\boldsymbol{0.724}\pm0.014$ & $0.718 \pm 0.033$ & $0.718 \pm 0.024$ \\\hline
\end{tabular}
}
\captionsetup{font=scriptsize}
\caption{Comparison of weighted F1 scores.}\label{accuracy}
\end{table}

\begin{figure}[t]
\centering

\begin{tikzpicture}[scale=0.5,>=latex]

\begin{axis}[
 title={Runtime vs Overlapping samples},
    xlabel={Overlapping Samples [Total=100]},
    ylabel={Time[ms]},
    xmin=0, xmax=100,
    ymin=175, ymax=225,
    xtick={0,20,40,60,80,100},
    ytick={180,190,200,210,220},
    legend pos=north west,
    ymajorgrids=true,
    grid style=dashed,
]
 
 \addplot[
    color=orange,
    mark=square,
    ]
    coordinates {
    (20,179)(40,187)(60,200)(80,215)
    };
    \addlegendentry{features$_T$=5} 
\addplot[
    color=blue,
    mark=square,
    ]
    coordinates {
    (20,182)(40,189)(60,203)(80,217)
    };
    \addlegendentry{features$_T$=10}
\addplot[
    color=red,
    mark=square,
    ]
    coordinates {
    (20,184)(40,190)(60,205)(80,220)
    };
    \addlegendentry{features$_T$=20}

\end{axis}

\begin{scope}[xshift={8.1cm}]
\begin{axis}[
    title={(b) Runtime vs d},
    xlabel={d [Total=20]},
    ylabel={Time [ms]},
    xmin=0, xmax=20,
    ymin=50, ymax=71,
    xtick={0,5,10,15,20},
    ytick={50,54,58,62},
    legend pos=north west,
    ymajorgrids=true,
    grid style=dashed,
]
 
 \addplot[
    color=orange,
    mark=square,
    ]
    coordinates {
    (5,52)(10,55)(15,60)
    };
    \addlegendentry{Overlapping samples = 5} 
\addplot[
    color=blue,
    mark=square,
    ]
    coordinates {
    (5,53)(10,56.5)(15,62)
    };
    \addlegendentry{Overlapping samples = 10}
\addplot[
    color=red,
    mark=square,
    ]
    coordinates {
    (5,56)(10,59)(15,64)
    };
    \addlegendentry{Overlapping Samples = 15}
\end{axis}
\end{scope}
\end{tikzpicture}
\captionsetup{font=footnotesize}
\caption{Scalability in Malicious Setting}
\label{fig:malicious}
\end{figure}
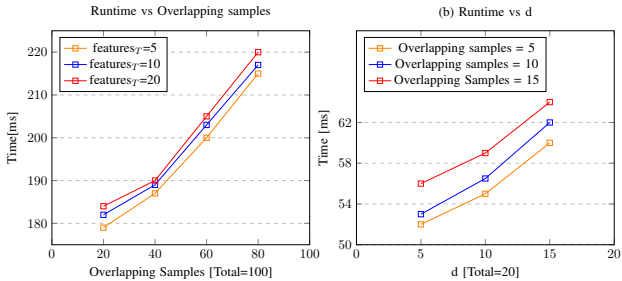

As compared to other privacy preserving work on ML, secure FTL training highly benefits due to local evaluation of the neural network, as:
    \begin{enumerate}
    \item Due to the restrictions on division and exponentiation, many MPC-based solutions require approximation of the non-linear activation function for each layer of the model in each iteration, leading to loss of accuracy. For instance, SecureML incurs a 1\% loss in the evaluation of a 2 layer Fully-Connected Neural Network with 128 neurons in each layer.
    Since in our case, the model evaluation is local and approximations are made for evaluation of a comparatively low depth circuit once after each iteration, higher level of accuracy is observed. 
    \item ML platforms like Tensorflow can be used for the evaluation. In contrast to this, most MPC-based solutions for secure self learning models, including use of MP-SPDZ Framework, are not compatible with such ML platforms.
    \item Security needs to be adapted for low depth circuits of gradients and loss calculation, yielding highly practical implementations even with malicious security. 
    \end{enumerate}

\section{Conclusion}\label{future} In this paper, we establish the practicality and scalability of secure FTL for both semi-honest and malicious setting. Our techniques for training bring multifold improvement in the run-time and communication cost as compared to the previous on HE-based approach. Given the lack of actively secure machine learning protocols, we hope this paper would pave the way for future works that guarantee strong security. 

\subsection{Future Work}
For the actively secure case, in contrast to our work over fields i.e. modulo prime, another direction can be an implementation over rings of the form $2^k$ using SPDZ-2k protocol \cite{SPDZ2k}. Such an setting can leverage the efficiency of native operations in a 32-bit/64-bit standard CPU to show efficient results. \\
While the security we introduce for collaborative calculation of gradients will suffice for many real-life scenarios, some applications might require a stronger security guarantee. Since trained model parameters are a function of training data, in some cases they can reveal private information about the datasets involved in training \cite{differential}. For security against such attacks, our techniques can be merged with secure aggregation techniques introduced in \cite{google_aggregate}. As secure aggregation involves revealing the aggregation of trained parameters from multiple parties for updation of a shared global model, SPDZ remains a sound option for a FTL model involving more than two parties and secure aggregation, because of its efficient scalability. The online phase of SPDZ scales linearly as the number of participants are increased, thus leading to an even more robust yet practical system. Apart from this, guarding of revealed gradients in the current model with differential privacy \cite{dp} can be a considered as a solution.

\bibliographystyle{IEEEtran}
\bibliography{Bigdata}

\end{document}